\documentclass{aa}
\usepackage{astron} 
\usepackage{float,epsfig,psfig}
\usepackage{pstricks}
\begin{document}

\title{XMM-Newton observations of the Lockman Hole: III. A relativistic 
Fe line in the mean X-ray spectra of type-1 and type-2
AGN \thanks{Based on observations obtained with {\em XMM-Newton}, an ESA
science mission with instruments and contributions directly funded by ESA
Member States and the USA (NASA)}}

\author {A. Streblyanska\inst{1}
                \and G. Hasinger\inst{1}
              \and A. Finoguenov\inst{1}
              \and X. Barcons\inst{2}
               \and  S. Mateos\inst{2}
               \and  A. C. Fabian\inst{3}}
 \offprints{A. Streblyanska,\\ \email{alina@mpe.mpg.de}}

   \institute{
   Max-Planck-Institut f\"ur Extraterrestrische Physik, Giessenbachstra\ss e,
 D-85740,  Garching, Germany
\and Instituto de F\'{\i}sica de Cantabria (CSIC-UC), 39005 Santander, Spain
\and Institute of Astronomy, Madingley Road, Cambridge CB3 0HA}

   \date{Received 
%8 May 2004; 
/ accepted ... }

\authorrunning{Streblyanska et al.}
\titlerunning{Fe line in type-1/2 AGNs }

\abstract{Using the 770 ksec XMM-Newton survey of the Lockman Hole field in
combination with extensive optical identifications of the AGN population, we
derive an average rest-frame spectrum of AGN types-1 and 2. The most
prominent feature in the averaged spectrum is a strong fluorescent
Fe line. In both type-1 and type-2 AGN, a clear relativistic line profile is revealed. A laor line profile with an inner disk radius smaller than the last stable orbit of a Schwarzschild black hole is most consistent with the data, indicating that the average supermassive black hole has significant spin. Equivalent widths of the broad relativistic lines range between 400--600 eV. We used the disk reflection model to compare the observed
strength of the line with  the amplitude of the reflection
component, concluding that to consistently describe the observations
the average iron abundance should be about three times the solar value.
  \keywords{Surveys -- Galaxies: active -- {\itshape
(Galaxies:)} quasars: general -- X-rays: general -- galaxies: nuclei} }

\maketitle

\section{Introduction}

The X-ray background is the echo of the growth of the Black Holes, which we
see in the center of most galaxies today. A large fraction of the background
has been resolved into discrete sources, being almost 100\% below 2 keV and 
about
50\% at 10 keV (Worsley et al. 2004). Since the peak of the background is at
30 keV, much still has to be resolved and we need detailed population
synthesis models to understand the total background. A significant
uncertainty still exists in the total flux of the XRB (Revnivtsev et
al. 2003).

The observed background spectrum can be reproduced well by models folding
the observed X-ray luminosity function and its evolution over cosmic time
with AGN spectra observed through different amounts of neutral hydrogen
column density following unified AGN models (Comastri et al. 1995, Gilli
et al. 2001).

An important ingredient to the background synthesis models is the average
X-ray spectrum of AGN, which at the moment has been largely taken from local
samples of Seyfert-1 and Seyfert-2 galaxies. The typical model includes at
least two parameters: the slope of the AGN spectrum and absorbing column
(Mainieri et al. 2002).  Here we explore the role of the fluorescent Fe
line. A possibility to detect a feature in the cosmic spectrum arising from
the Fe line has been first studied by Schwartz (1988), where narrow Fe lines
were assumed. However, a diversity in line shapes have been recently
reported, as relativistic iron lines have been discovered in some objects
with ASCA (Tanaka et al. 1995) and confirmed with XMM-Newton (Wilms et
al. 2001, Fabian et al. 2002). The most recent evidence is
controversial. Only a small fraction of Seyfert galaxies seems to emit
relativistic iron lines (Boller et al., in prep., Balestra et al. 2004) and the
line shapes observed with XMM-Newton and Chandra are more
complicated. Frequently a narrow component is observed in the Fe line and
the analysis of the broad, relativistic component is often hampered by
systematic uncertainties in modelling the continuum.

In this paper we have used the longest and most sensitive observation of the
X-ray background by XMM-Newton, a 770 ksec observation in the Lockman Hole
(Hasinger 2004). The fraction of optically identified sources with redshifts
and a sufficient number of X-ray photons is quite high in this field, so that 
for
the first time we are able to analyse the mean rest-frame spectra for a
representative sample of type-1 and type-2 AGN at considerable redshifts. We
find significant features in the average rest frame spectra: a broad line,
with a peak at $\sim 6.4$ keV for the mean spectrum of type-1 AGN and 
a broad line, possible with a superposed narrow line with a peak at 6.4 keV 
for the mean type-2
spectrum. We model the broad component as a relativistically broadened
fluorescent Fe line from the accretion disk around the Kerr
(rotating) black hole (Laor 1991) with an equivalent width of $\sim 560$ eV
and $\sim 460$ eV for the average spectrum of type 1 and 2 AGN,
respectively.

The paper is structured as follows: \S 2 describes the observations, \S 3
the results of spectral analysis, with a discussion of the results presented
in \S 4. We summarize in \S 5.

\section{XMM-Newton observations and data reduction}

The X-ray results reported in this paper are obtained from XMM-Newton
observations of the Lockman Hole field, a special region in the sky with a
extremely low Galactic hydrogen column density, $N_{\rm H} = 5.7 \times
10^{19}$ cm$^{-2}$ (Lockman et al. 1986).

The field was centered on the sky position RA 10:52:43 and DEC
+57:28:48 (J2000) and observed during the three periods in 2000, 2001 and
2002 (PV, AO-1 and AO-2 phases). The remaining exposure time after cleaning from high background periods is approximately 770 ks. The PN and MOS data were preprocessed by the $XMM$ $Survey$ $Science$ $Centre$ with the XMM standard Science Analysis System (SAS, version 5.3.3) routines, using the latest calibration data. We have derived the event files from the 16 individual observations of the Lockman Hole in the standard way (see Worsley et al. 2004).

Source and background spectra and the response matrices, {\tt arf} and 
{\tt rmf}, have
been extracted for each source and each individual observation. The {\it evselect} tool is used to extract the spectrum and background region, which is defined as an annulus around the source, after masking out nearby sources. MOS1 plus MOS2 and independently pn source and background spectra and response matrices
were combined for each source. Before spectral fitting, all spectra are
binned with a minimum of 30 counts per bin in order to be able to apply the
$\chi^2$ minimization technique. In this process, the background count rate
is rescaled with the ratio of the source and background areas.

\subsection{X-ray sample}

Images, background maps and exposure maps were generated for each
observation and detector individually in the standard 5 energy bands:
0.2--0.5, 0.5--2, 2--4.5, 4.5--7.5 and 7.5--12 keV. We ran the SAS source
detection algorithm eboxdetect-emldetect on each instrument independently,
using the five energy bands simultaneously. Afterwards we cross-correlated 
the pn
source list with the lists obtained for MOS1 and MOS2 to build a complete
unique catalogue of sources. As a last step we also cross-check our catalogue
with the source catalogue by Mainieri et al. \cite{mai02} from the PV
observation.

We selected a sample of 104 X-ray sources spectroscopically identified by
Lehmann et al. (2001) and some additional spectroscopic redshifts for new XMM-Newton sources obtained at the Keck telescope in 2003 (Lehmann et al. \cite{lehmann05}) with counts larger than 200 in the 0.2--10 keV band.
The bulk of the source identification comes from the flux limited ROSAT
0.5--2 keV sample.  Most of these sources have been analysed individually
by Mainieri et al. \cite{mai02}. Mateos et al. \cite{mat05} present a detailed individual spectral analysis of the brightest sources, using the same data
utilized in this paper. 

In the following we will refer to type-1 and type-2 objects using the
optical spectroscopic classification (see Schmidt et al. 1998, Lehmann et
al. 2001). Our final sample includes 53 type-1 AGN, 41 type-2 AGN and 10
galaxies.  Throughout the paper, errors correspond to the 90\% confidence level
for one interesting parameter ($\Delta \chi^2=2.706$).

\begin{figure} \centering
 \includegraphics[width=8.3cm]{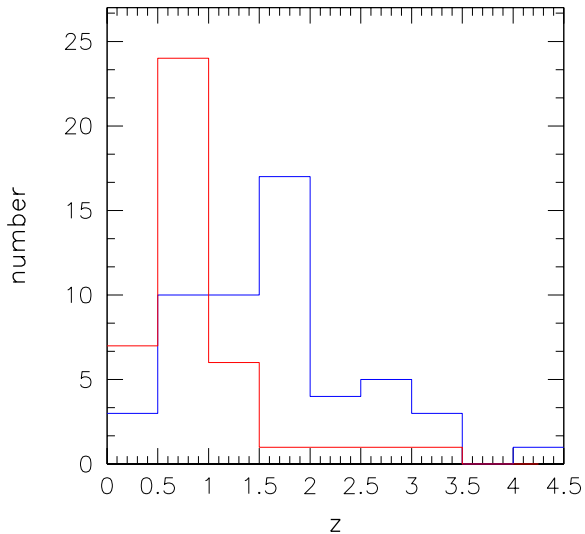}
   \caption{Redshift distribution of the LH X-ray sources in our selected
     sample. The blue and red lines correspond to type-1 and type-2 AGN,
     respectively. Among type-2 AGNs, the objects at low redshifts ($z<1$)
     constitute 57\%.}
  \label{fig1}
 \end{figure}

\section{Results}
\subsection{Spectral analysis}
The redshift distribution of the sources in our sample in shown in
Fig.\ref{fig1} and peaks at redshifts below 1 for type-2 AGN and close to
2 for type-1 AGN.

We used {\it XSPEC (v11.2)} for the spectral fitting analysis. All spectra have
been extracted in the 0.2--10 keV band, MOS1 and MOS2 spectra have been
merged. For every source we use a separate response matrix and ancillary
response files, where a change in the filter between the observations is
propagated as described in Mateos et al. \cite{mat05}.

For every object in our sample we fit the spectrum with a single power law
model modified by an intrinsic absorption, having the slope of the power
law, its normalization and the absorbing column as free parameters. We save
the ratio of the data to the model as well as a reconstructed unfolded
spectrum. The next step consists of shifting the spectrum to the rest
frame. For the ratio we simply increase the energies by the factor of
$(1+z)$, while for the unfolded spectra, we use the 2--10 keV rest frame
energy band to renormalize the spectra to the same value for further
averaging.

As the redshifts of the sources vary, their rest frame spectra have
different energy boundaries. To provide an average spectrum, we select a
common energy grid. We used a bin width of 0.25 keV for energies lower than
8 keV and a bin width 2 keV above. Such a choice is determined by the available counts. In a separate analysis, to study the Fe line profile in more detail, we rebinned the data using a binsize of 0.23 keV between 4 and 7 keV for the type-1 AGN spectra. \\
As the input spectra often have energy bins spread over two or more energy
bins of the output spectrum, in such instance we use the Monte-Carlo method
to choose the bin into which to add the data. As a final operation, we
normalize the flux in each energy bin of the output spectrum by the number
of points that were added to it. We propagate the errors accordingly.

Such an averaging is similar between the ratio and the unfolded spectra and
both methods yield comparable results. As long as the redshifts of the objects in our sample are of order 1, it is possible to observe the rest frame averaged spectrum above 12 keV. 

The averaged spectra for both type-1 and type-2 AGN show a prominent spectral feature at 6.4 keV. We observed an excess in the data/model ratio between
4--7 keV for the both type-1 and type-2 AGN (Fig. 2 and Fig. 3). 

Such a feature can be due either to a broad emission line or to complex 
absorption. For the latter we have modelled the spectrum in terms of a 
partially-covered source (e.g. Boller et al. 2003) and obtain a good fit for the following parameters: $\Gamma = 2_{-0.05}^{+0.06}$, $N_{\rm H} = 14^{+2}_{-2} \times 10^{22}$ cm$^{-2}$ and covering fraction $= 0.4_{-0.05}^{+0.05}$. 
%Such strong partial-covering is unlike absorption seen in nearby sources an would require them to be strongly similar. 
Such a strong partial-covering is usually not seen in nearby sources and to observe it in the average spectrum would require the individual sources to be very similar to each other. A variability analysis of the individual source components may be a useful test. In the remainder of this paper we explore the broad emission line interpretation.

%(see Sect. \ref{sec:discuss})

\begin{figure} \centering
\includegraphics[width=7.cm, angle=270]{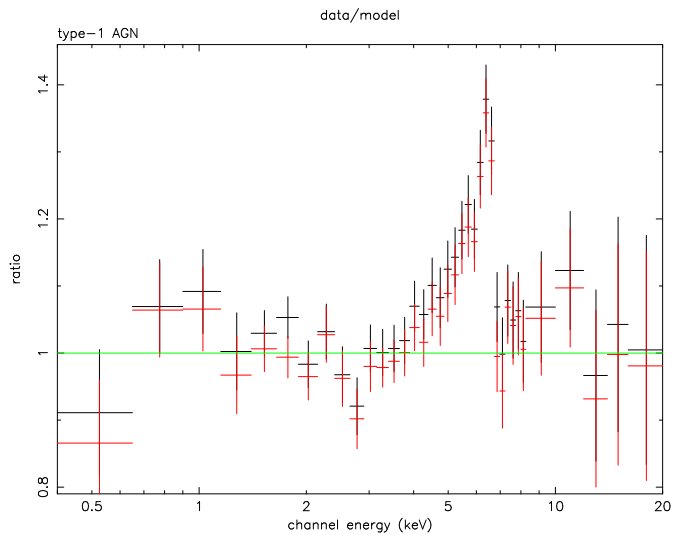}
 \caption{The resulting ratio plots from mean unfolded spectra type-1 AGN spectra obtained by using averaging methods. The unfolded spectra are fit with a simple power-law model in the ranges 0.2--3 and 8--20 keV (excluding the portion
   of the spectrum where the emission associated with the relativistic Fe
   line is present). The EPIC-pn (black) and MOS (red) spectral
   data show a clear broad excess at 4-8 keV in the rest-frame. } 
\label{fig2}
\end{figure}

\begin{figure} \centering
 \includegraphics[width=7.cm, angle=270]{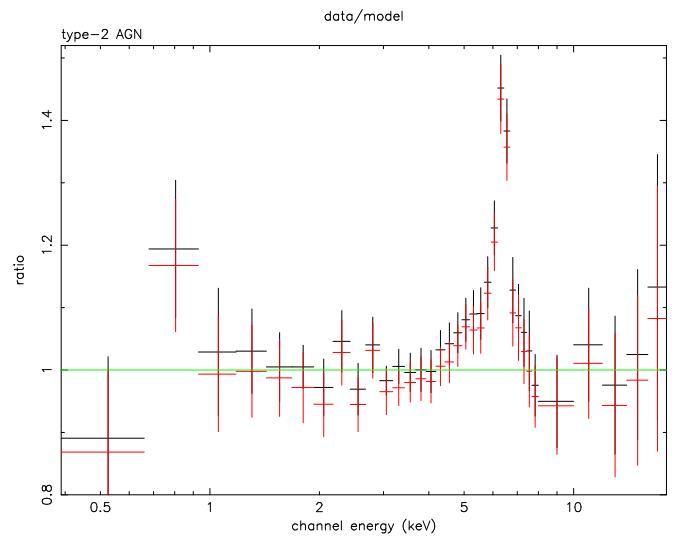}
   \caption{The 0.2--3 and 8--20 keV power-law model fits for the PN (black),
     MOS (red) spectral data, showing a broad excess similar to
     type-1 AGN but in conjunction with a narrow component near 6.4
     keV.} 
  \label{fig3}
 \end{figure}

%Moreover we detected a strong soft component (0.4--2 keV) for
%type-2 AGN in both pn and MOS data (Fig.3). 

\subsection{Origin of the broad line feature in the type-1 AGN spectrum}
\label{line}

\begin{figure*}
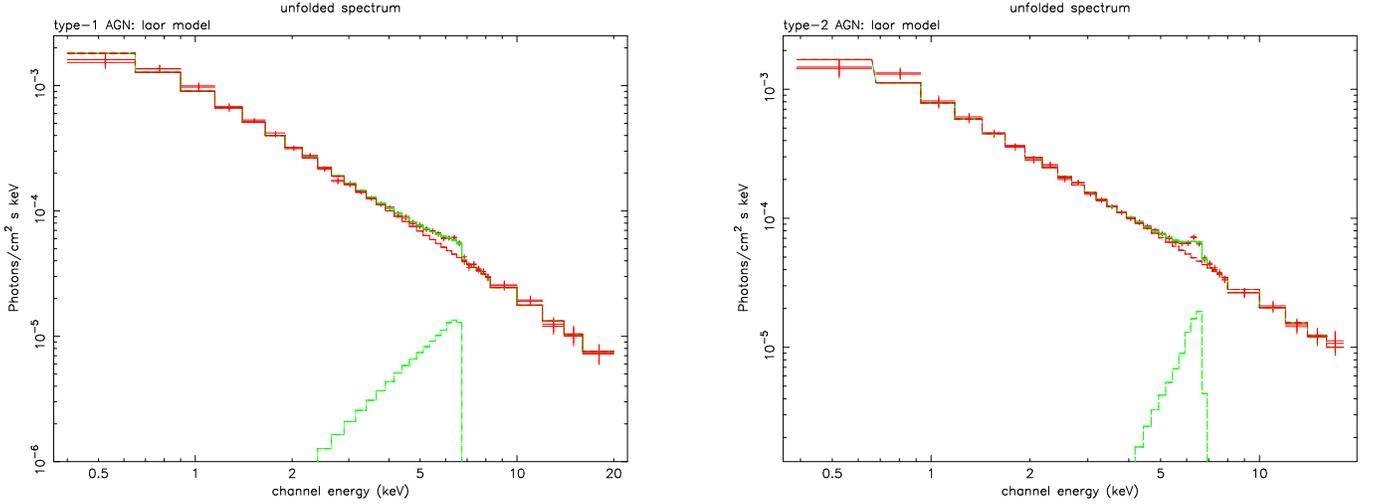

\parbox{8.3cm}{\resizebox{\hsize}{!}{\includegraphics[angle=270]{uf_agn1_laor.ps}}}
\hfill
\parbox{8.3cm}{\resizebox{\hsize}{!}{\includegraphics[angle=270]{uf_agn2_laor.ps}}}
 \caption{The unfolded stacked X-ray spectra and best fit model (laor) representative of the type-1 and type-2 AGN.}
\label{fig4}
\end{figure*}

The energy of this broad line indicates that Fe atoms responsible for the line are in a low ionization state, with mean energy near 6.4 keV
(in the rest-frame). Most probably such a broad and intense Fe $K_{\alpha}$
line is due to low (moderate) ionization states of iron (i.e., $<$Fe XVI).
The line shape appears to be as predicted from disk theory and skewed toward
energies lower than the rest energy of the emission line. The simplest
interpretation of the skewed shape of the line is that it is due to Doppler
and gravitational redshifts from the deep gravitational potential (the inner
parts of a disc around a massive black hole). A strong red wing is indicative
of gravitational redshifts close to a central black hole, and accretion disk
models provide an excellent description of the data. Such line profile can
be fitted with a relativistic profile from an accretion disc around either a
Schwarzschild (non-rotating) or a Kerr (rotating) black hole ({\it XSPEC} models
{\tt diskline} and {\tt laor}, respectively; Fabian et al. 1989; Laor 1991).

In both models we fixed the line energy to 6.4 keV and the emissivity index 
(-2 for {\tt diskline} and 3 for {\tt laor}) and fitted
the rest of parameters: inner $R_{in}$ and outer $R_{out}$ radii and
inclination angle of the disc $i$ (see Table.~\ref{t:fit}). The results for the inclination angle are close to the standard value of 30 deg observed for a wide sample of AGN, consistent with the sharp drop near 7 keV and observed peak around the rest energy (6.4 keV). 

The {\tt laor} and {\tt diskline} model yield similar reduced $\chi_\nu^2$, but the disk line model, at least for type-1 AGN spectra prefers an inner disk radius at the last stable orbit and an unreasonably small outer radius.
%$R_{out}=28 R_{s}$ (where $R_{s} = 2GM/c^{2}$ is the Schwarzschild radius
%for mass M). With fixed $R_{out}=1000$ we obtained $\chi_\nu^2 = 1.58$ and
%EW = 280 eV. The trend toward smaller resulting parameters implies that an emission region must to be close to the black hole.

In order to estimate the errors for the equivalent widths, we assumed the relative errors to be the same as that of the line flux, because the errors in the continuum at the position of the Fe line position are negligible compared to this. It was found to have an average equivalent width of $560_{-110}^{+90}$ eV ({\tt laor} model).

\begin{table*} { \begin{center} \footnotesize
  {\renewcommand{\arraystretch}{1.3} \caption[]{
Results of the spectral fitting in 0.4-20 keV band with a four-component model consisting of a power law and a line model with two absorption components, one representing intrinsic absorption (a free parameter during the fit) and the other was fixed at the Galactic column density of $5.7 \times 10^{19}$ cm$^{-2}$ ( $wabs\ast wabs(po+linemodel))$. The line models are Gaussian, diskline and laor.  From our model fits, we computed the slope of a power law spectrum (photon index $\Gamma$), a intrinsic column density $N_{\rm H}$ and the line parameters.
%In the bracket show number if we fit in 0.4-10 keV band. 
%All models give us relatively small $\chi^2$ due to statistical dependence of the data.
 }  \label{t:fit}}
  \begin{tabular}{lllllllllll}
\hline   
\hline     
 Model & &  & &  \multicolumn{2}{c}{parameter}    \\
\hline  
&&&&Type-1 AGN\\
\hline  
& $\Gamma$ & $N_{\rm H}$ & Energy & EW & $\sigma$ & $R_{in}$ & $R_{out}$ & $\beta$ & $i$ & $\chi_\nu^2/dof$ \\
&  &  cm$^{-2}$ & keV & eV & keV & $R_{g}$ & $R_{g}$ &  & deg &  \\
\hline  
 Gaussian & $1.75_{-0.02}^{+0.02}$ &  $7.9^{+0.9}_{-0.6} \times 10^{20}$ & $5.9^{+0.2}_{-0.1}$ & $420^{+35}_{-30}$ & $0.69_{-0.05}^{+0.05}$ & $ - $ & $ - $ & $ - $ & $ - $  & $1.35/68 $ \\
diskline & $1.73_{-0.01}^{+0.02}$ &  $7.5^{+1}_{-1} \times 10^{20}$ & $6.40^*$ & $480^{+60}_{-60}$ & $ - $ & $6.4^{+1.8}_{-0.4}$ & $22_{-3}^{+6}$  & $2^*$ & $29.1_{-0.9}^{+3.2}$  & $0.70/67$ \\
laor & $1.74_{-0.02}^{+0.01}$ & $7.5^{+1.0}_{-0.9} \times 10^{20}$ & $6.40^*$ & $560^{+90}_{-110}$ & $ - $ & $3.1^{+1.2}_{-0.8}$ & $ 400_{-120}$ & $3^*$ & $29.6_{-1.7}^{+1.8}$ & $0.73/67 $        \\
\hline  
\hline  
&&&&Type-2 AGN\\
\hline  
& $\Gamma$ & $N_{\rm H}$ & Energy & EW & $\sigma$ & $R_{in}$ & $R_{out}$ & $\beta$ & $i$ & $\chi_\nu^2/dof$ \\
&  & cm$^{-2}$ & keV & eV & keV & $R_{g}$ & $R_{g}$ &  & deg &  \\
\hline  
Gaussian & $1.59_{-0.01}^{+0.02}$  &  $5.1^{+0.9}_{-0.8} \times 10^{20}$ & $6.34_{-0.03}^{+0.03}$ & $280^{+25}_{-15}$ & $0.27_{-0.02}^{+0.04}$ & $ - $ & $ - $ & $ - $ & $ - $  & $0.83/64$ \\
diskline & $1.59_{-0.01}^{+0.01}$ & $5.0^{+0.8}_{-0.9} \times 10^{20}$ &  $6.40^*$ & $320^{+25}_{-30}$ & $ - $ & $6.0^{+0.1}$ & $800_{-360}^{+400}$  & $2^*$ & $33.2_{-2.2}^{+1.4}$  & $0.80/63$ \\
laor & $1.61_{-0.01}^{+0.02}$  & $5.3^{+0.9}_{-0.8} \times 10^{20}$ & $6.40^*$ & $455^{+45}_{-30}$ & $ - $ & $6.2^{+1.5}_{-0.2}$ & $ 400_{-100}$ & $3^*$ & $30.1_{-1.8}^{+0.6}$  & $0.66/63$ \\
\hline          
\hline  
  \end{tabular}
  \end{center}
\hspace*{0.3cm}{\footnotesize}
$^*$ fixed 
}
  \end{table*}

\subsection{Type-2 AGN}

Our initial fit for the unfolded spectrum was an absorbed power law with a
Gaussian component to represent the iron line. Moreover we used, as for type-1 AGN spectra, disk line models to fit our asymmetric line profiles. Disk line models provide a good fit to the red wing of the data. Results are shown in Table.~\ref{t:fit}.  As an additional test, we use an additional zero-width Gaussian to fit a possible narrow line component in the Fe emission separately, which could come from reflection in material much further away from the accretion disk, e.g. in the torus, but this does not improve our $\chi_\nu^2$ significantly and therefore we neglect this Gaussian.

The unfolded spectrum for the type-2 AGN with skewed line wings looks similar to that of the type-1 AGN, except that an additional narrow iron line component at a rest energy $\sim 6.4$ keV seems to be present in the type-2 AGN. 
%which is remarkable close to the Fe I at 6.40 keV). 
%Such sharp Fe $K_{\alpha}$ line reemitted from cold matter and must originate far from the inner accretion disc, perhaps in the putative molecular torus (NO!). 

Again, we assumed the relative error of the equivalent width to be that of the line flux. It was found to have an average equivalent width of $455^{+45}_{-30}$ eV (laor model).

\section{Discussion}
\label{sec:discuss}

A Doppler-broadened line with a full width at half-maximum, $FWHM \sim
150,000$ km s$^{-1}$ can only be produced in the inner region of a
relativistic accretion flow by fluorescence (Fabian et al. 2002). The observed FWHM and EW is
larger than the average value found in Seyfert 1 galaxies ($\sim$ 500 eV,
e.g. Nandra et al. 1997), but similar to the FWHM of some of the brightest nearby active galaxies (for example, to MCG-6-30-15 with z=0.007749). 

One of the explanations for the large EWs for these objects can be given by a model including ionized disc reflection with lines and edges from different ionization stages of iron blurred together by relativistic effects (Ross \& Fabian \cite{ros93}, Fabian et al. 2002b). In this model, the value of the Fe abundance determines the relative 
importance of the line complex in respect to the comptonization bump,
seen at energies exceeding 10 keV. Since the observed equivalent width of the 
broad Fe line is high, while the amplitude of a possible reflection component observed above 10 keV is small,
this model yields a high Fe abundance as an explanation of the observation.
This is in fact a novel method of measuring Fe abundance in QSO at X-rays,
which in addition to the determination of redshifts will allow future X-ray
surveys to give insight in the chemical evolution of the host environment of QSO.  This model has been applied for several nearby objects and can account for all the observed spectral features, but yields a large iron overabundance of 3-7 $\times$  solar (Tanaka et al. 2004, Boller et al. 2003). In our sample we also most likely obtain the large EW as the result of a high metallicity. To test this idea, we used the blurred ionised reflection models 
(Ross \& Fabian \cite{ros93})
with an iron abundance of 3 $\times$ solar (as needed for MCG-6-30-15) in order to obtain a high EW. The results describe our data very well. 

An additional reason why our large equivalent width may be connected with a large metallicity is that most of our objects are quite distant and rather luminous AGNs.  As shown by Shemmer et al. (2004), the metallicity is correlated with the accretion rate, which is in turn related to the luminosity.  It is possible that the large equivalent widths are representative of a high metallicity, since these objects are both distant and luminous.

Our results are most consistent with a Laor model with inner disk radius well inside the maximum stable orbit for a Schwarzschild black hole, implying that the average  X-ray bright AGN should contain rotating Kerr BH at their centres  \footnote{ The Laor (1991) model which is appropriate for 
spin a/m=0.998 gives  a best fit $R_{\rm in}=3$. Dov\v{c}iak, Karas \& 
Yaqoob (2003) plot the maximum redshift as a function of radius and spin 
(their Fig.~2). Assuming that the spectral fit is determined by the 
maximum redshift and that the emitting region is outside the radius of 
marginal stability (see Krolik \& Hawley 2003 for a discussion of that 
point), $R_{\rm in}=3$ corresponds to $a/m> 0.6$.}. If this is the case, it implies that the spin will affect demographic arguments relating the AGN luminosity function to the mass function of remnant black holes in nearby glaxies (So\l tan 1982) since the radiative efficiency of accretion will increase. Indeed, this is postulated by recent work updating the So\l tan argument with modern galaxy -- black hole correlation functions and AGN energetics based on hard X-ray surveys (Yu \& Tremaine \cite{yutr02}, Elvis, Risaliti \& Zamorani \cite{elv02}, Marconi et al. \cite{mar04}).Accretion is expected to build massive black holes  with significant spin (Hughes \& Blandford \cite{hug03}; Volonteri et al. 2004).

%contain rotation Kerr BH at their centres  
%If $R_{\rm in}\sim 3R_{\rm g}$ is the typical innermost stable orbit (see
%Krolik \& Hawley \cite{kro02} for discussion on this point) then the
%spin parameter $a/m\sim 0.8$ (Bardeen et al. \cite{bar72}.
%If this is the case, it implies that the spin will affect demographic arguments relating the AGN luminosity function to the mass function of remnant black holes in nearby glaxies (So\l tan 1982) since the radiative efficiency of accretion will increase. Indeed, this is postulated by recent work updating the So\l tan argument with modern galaxy -- black hole correlation functions and AGN energetics based on hard X-ray surveys (Yu \& Tremaine \cite{yutr02}, Elvis, Risaliti \& Zamorani \cite{elv02}, Marconi et al. \cite{mar04}). Our result is also relevant to the merger history of BH (Hughes \& Blandford \cite{hug03}). 

The intensity of the relativistic line in our mean spectra implies that
this line must be present in the majority of AGN in our sample. Using a similar
technique we followed the same procedure with a source list and spectra from 
the XMM-Newton observation of the Chandra Deep Field South (Streblyanska et al. 2004). The results are consistent with our analysis of the LH and 
also with an independent analysis of the Chandra deep field data by Brusa et al. \cite{bru05}. 
We also investigated, whether the signal could be dominated by a small number of objects with extremely bright features. For this purpose we applied a $\sigma$-clipping before averaging our spectral bins, but the resulting spectra do not differ significantly.

In future work we plan to investigate the dependence of the line on other parameters, such as redshift or luminosity. We will also consider a composite model with a range of inclinations.

\section{Conclusions}
This work includes analysis of EPIC-pn and MOS1/2 data.  We present results
of the X-ray spectral analysis of the 770 ksec deep survey observations
obtained with XMM-Newton on the Lockman Hole. We have derived the average X-ray
spectral properties of sources spectroscopically identified with the
Keck telescopes.  With data from the EPIC-pn and MOS detectors, we
used a sample of 103 sources with redshift identification and photon
counts $> 200$.  We used an averaging technique to determine average spectra
of the different AGN populations.  We parameterize the spectrum with a simple power law model. The continuum at 4--7 keV shows residuals above the power-law
model, suggesting the presence of a strong broad feature, most likely 
a relativistic iron disk line.
In the mean spectra of type-1 and type-2 AGN  we find evidence for a broad line at rest-frame energy $\sim 6.4$ keV with EW $\sim 560$ eV
and $\sim 460$ eV (laor model), respectively. We interpret this line as the fluorescent Fe $K_{\alpha}$ line commonly observed in Seyfert galaxies.  Our mean spectrum for the type-2 AGN indicates in addition to the broad line a narrow component.

The most interesting conclusions which follow from our analysis are:

The average rest-frame spectrum of the XRB sources shows a strong, relativistic iron line, possibly due to a high metallicity in the average population. 
The strong red wing of the line feature indicates some component of
spin in the average black hole. 

A very broad line feature is expected in the average X-ray background spectra (Fabian et al. 2000, Gilli et al. 2001), which should be included in future population synthesis models for the X-ray background.

In spite of the variety and complexity of Fe features at low z bright
objects (both type-1 and type-2 Seyfert), the average line shape at 
higher redshift is consistent with a relativistic disk line  
for type-1 and type-2 AGN with some evidence of an additional narrow 
component for type-2 AGN. 

These findings strongly reinforce the science case of the XEUS/Constellation-X mission; with the very large collecting area we can expect to measure 
the relativistic Fe line parameters of individual objects out to very high 
redshifts (z $\approx$ 10) and explore the evolution of black hole spin.

\begin{acknowledgements}
We thank T. Boller for helpful comments and discussions that improved the manuscript. This work was supported by the German \emph{Deut\-sches Zentrum f\"ur Luft- und Raumfahrt, DLR\/} project number 50 OR 0207 (A. Finoguenov) and the Max-Planck society, MPG. XB acknowledges financial support from the Spanish Ministerio de Educaci\'on y Ciencia, under project ESP2003-00812.  

\end{acknowledgements}

\end{document}